\documentclass[aps,prl,twocolumn,showpacs,superscriptaddress,groupedaddress]{revtex4}

\usepackage{graphicx}  
\usepackage{dcolumn}   
\usepackage{bm}        
\usepackage{amssymb}   
\usepackage{amsmath}
\usepackage{braket}
\usepackage{soul}
\usepackage{gensymb}
\usepackage[usenames, dvipsnames]{color}
\usepackage[colorlinks = true,
            linkcolor = blue,
            urlcolor  = blue,
            citecolor = blue,
            anchorcolor = blue]{hyperref}

\begin{document}

\title{Berry curvature induced valley Hall effect in non-encapsulated hBN/Bilayer graphene heterostructure aligned with near-zero twist angle}
\author{Teppei Shintaku$^*$} \affiliation{Japan Advanced Institute of Science and Technology, 1-1 Asahidai, Nomi, 923-1292, Japan}
\author{Afsal Kareekunnan$^*$$^,$$^\dagger$}  \affiliation{Japan Advanced Institute of Science and Technology, 1-1 Asahidai, Nomi, 923-1292, Japan} 
\author{Masashi Akabori} \affiliation{Japan Advanced Institute of Science and Technology, 1-1 Asahidai, Nomi, 923-1292, Japan}
\author{Kenji Watanabe} \affiliation{National Institute for Materials Science, 1-1 Namiki, Tsukuba, 305-0044, Japan}
\author{Takashi Taniguchi} \affiliation{National Institute for Materials Science, 1-1 Namiki, Tsukuba, 305-0044, Japan}
\author{Hiroshi Mizuta} \affiliation{Japan Advanced Institute of Science and Technology, 1-1 Asahidai, Nomi, 923-1292, Japan}

\vskip 0.25cm
\date{\today}

\begin{abstract}
Valley Hall effect has been observed in asymmetric single-layer and bilayer graphene systems. In single-layer graphene systems, asymmetry is introduced by aligning graphene with hexagonal boron nitride (hBN) with a near-zero twist angle, breaking the sub-lattice symmetry. Although a similar approach has been used in bilayer graphene to break the layer symmetry and thereby observe the valley Hall effect, the bilayer graphene was sandwiched with hBN on both sides in those studies. This study looks at a much simpler, non-encapsulated structure where hBN is present only at the top of graphene. The crystallographic axes of both hBN and bilayer graphene are aligned. A clear signature of the valley Hall effect through non-local resistance measurement ($R_{\rm{NL}}$) was observed. The observed non-local resistance could be manipulated by applying a displacement field across the heterostructure. Furthermore, the electronic band structure and Berry curvature calculations validate the experimental observations.
\end{abstract}

\maketitle

\def\thefootnote{*}\footnotetext{These authors contributed equally to this work}\def\thefootnote{\arabic{footnote}}
\def\thefootnote{$\dagger$}\footnotetext{afsal@jaist.ac.jp}\def\thefootnote{\arabic{footnote}}

With the introduction of two-dimensional materials, the valley degree of freedom of carriers has gained much prominence in recent years \cite{Schaibley-2016, Vitale-2018, Xu-2014}. Materials like graphene and MoS$_{2}$ have two in-equivalent valleys at the K and K$'$ high symmetry points of their Brillouin zone, which can be interpreted as valley-up and valley-down, much like the spin degree of freedom of carriers. However, the fundamental criterion for a material to be valleytronic is to have a broken inversion symmetry. While single-layer graphene is symmetric, aligning graphene with hexagonal boron nitride (hBN) with a near-zero twist angle has proved to break the sub-lattice symmetry of the graphene layer \cite{Yankowitz-2012, Woods-2014, Song-2015, Yankowitz-2019}. Such a system has exhibited the valley Hall effect (VHE) due to the emergence of Berry curvature at the valley as a result of broken inversion symmetry \cite{Gorbachev-2014, Komatsu-2018, Li-2020}. As for bilayer graphene, the asymmetry can be introduced by either applying an out-of-plane electric field across the layers or by aligning with an hBN layer, both of which break the layer symmetry of the system as they introduce different potentials between the top and bottom layers of the bilayer \cite{McCann-2006, Min-2007, Oostinga-2007, Zhang-2009, Dean-2013}. Both methods have been employed to observe VHE in bilayer graphene in recent years \cite{Sui-2015, Shimazaki-2015, Endo-2019, Arrighi-2022}. 

In the case of single-layer graphene, it has been shown that both encapsulated and non-encapsulated graphene exhibit the valley Hall effect, provided either of the hBN (top or bottom) is oriented with graphene \cite{Gorbachev-2014}. As for bilayer graphene, the valley Hall effect is observed by hBN alignment only in encapsulated systems \cite{Endo-2019, Arrighi-2022}. Here hBN is present at the top and bottom of the bilayer graphene with one of the hBN aligned and the other misaligned by more than 10$^{\circ}$ to avoid the formation of a double moire pattern. However, it has been shown theoretically that aligning hBN with bilayer graphene in a non-encapsulated configuration can also break the symmetry of the system and induce Berry curvature \cite{Afsal-2020}. Considering the difficulty in fabricating encapsulated bilayer graphene heterostructure, in this study, we explore the VHE in non-encapsulated hBN/bilayer graphene heterostructure (hBN/bilayer graphene/SiO$_{2}$) with the hBN aligned with the bilayer graphene. We observed a strong VHE signal at the primary Dirac point through non-local electrical measurement. The VHE signal could be further manipulated by applying a displacement electric field across the layers. We also performed \textit{ab initio} calculations, which show that aligned hBN/bilayer graphene heterostructure has an intrinsic band gap and a non-zero Berry curvature. The band gap and the Berry curvature can be manipulated with an out-of-plane electric field applied across the layers.

\begin{figure*}[t]
\includegraphics[scale=0.6]{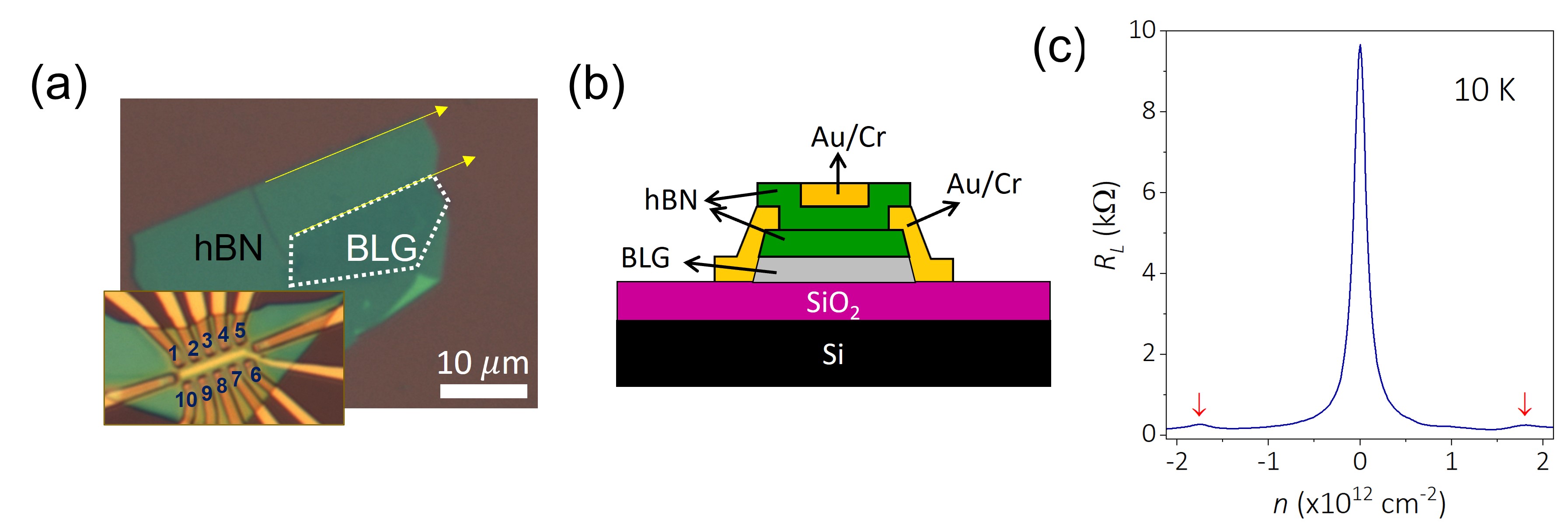}
\centering
\caption{\label{fig:one} (a) Optical image of hBN/bilayer graphene heterostructure aligned with near-zero twist angle. The dotted line shows the graphene region. The arrow indicates the edges of graphene and hBN, which are aligned. The inset shows the optical image of the final device. (b) The heterostructure schematic diagram shows the passivation hBN layer and the top gate electrode. (C) Measured gate characteristics of the device as a function of carrier density \textit{n}. The arrow indicates the two secondary Dirac peaks at the electron and hole sides. The measurement is performed at 10 K.}
\end{figure*}

The hBN/bilayer graphene heterostructure is fabricated following the dry transfer method \cite{Wang-2013}. Two devices denoted as Device A and Device B, are fabricated. The results presented in the main text are from Device A (see supplementary information for the results from Device B). Figure \ref{fig:one}a shows the optical image of the hBN/bilayer graphene heterostructure from which Device A is fabricated. The dotted line outlines the bilayer graphene area. The arrows indicate the edges of hBN and bilayer graphene, which are aligned with a near-zero twist angle. After etching the heterostructure into a Hall bar, edge contacts were fabricated \cite{Wang-2013}. Later a passivation hBN layer is transferred on top of the heterostructure, above which the top gate electrode is fabricated. Figure \ref{fig:one}a inset and Figure \ref{fig:one}b show the optical image and schematic diagram of the final device. Figure \ref{fig:one}c shows the gate characteristics of the device. The field-effect carrier mobility extracted from the linear region around the main Dirac peak (MDP) is $\approx$ 19 000 cm$^2$ V$^{-1}$ s$^{-1}$ and $\approx$ 26 000 cm$^2$ V$^{-1}$ s$^{-1}$ respectively for the holes and electrons. Apart from the MDP, secondary Diract peaks (SDP), which is the signature of the formation of moire superlattice, can be observed on both sides of the main Dirac peak. The SDP appears at $\pm$ 1.78 $\times$ 10$^{12}$ cm$^{-2}$, which corresponds to a moire superlattice period of 15.3 nm. This period is slightly larger than the maximum period (13.8 nm) expected for an hBN-graphene moire superlattice. The slight increase in the periodicity is attributed to the stretching of the bilayer graphene (supplementary information II) \cite{Woods-2014, Zihao-2019, Xingdan-2021}, which could be originated from the non-uniformity of the SiO$_{2}$ substrate.

\begin{figure*}[t]
\includegraphics[scale=0.6]{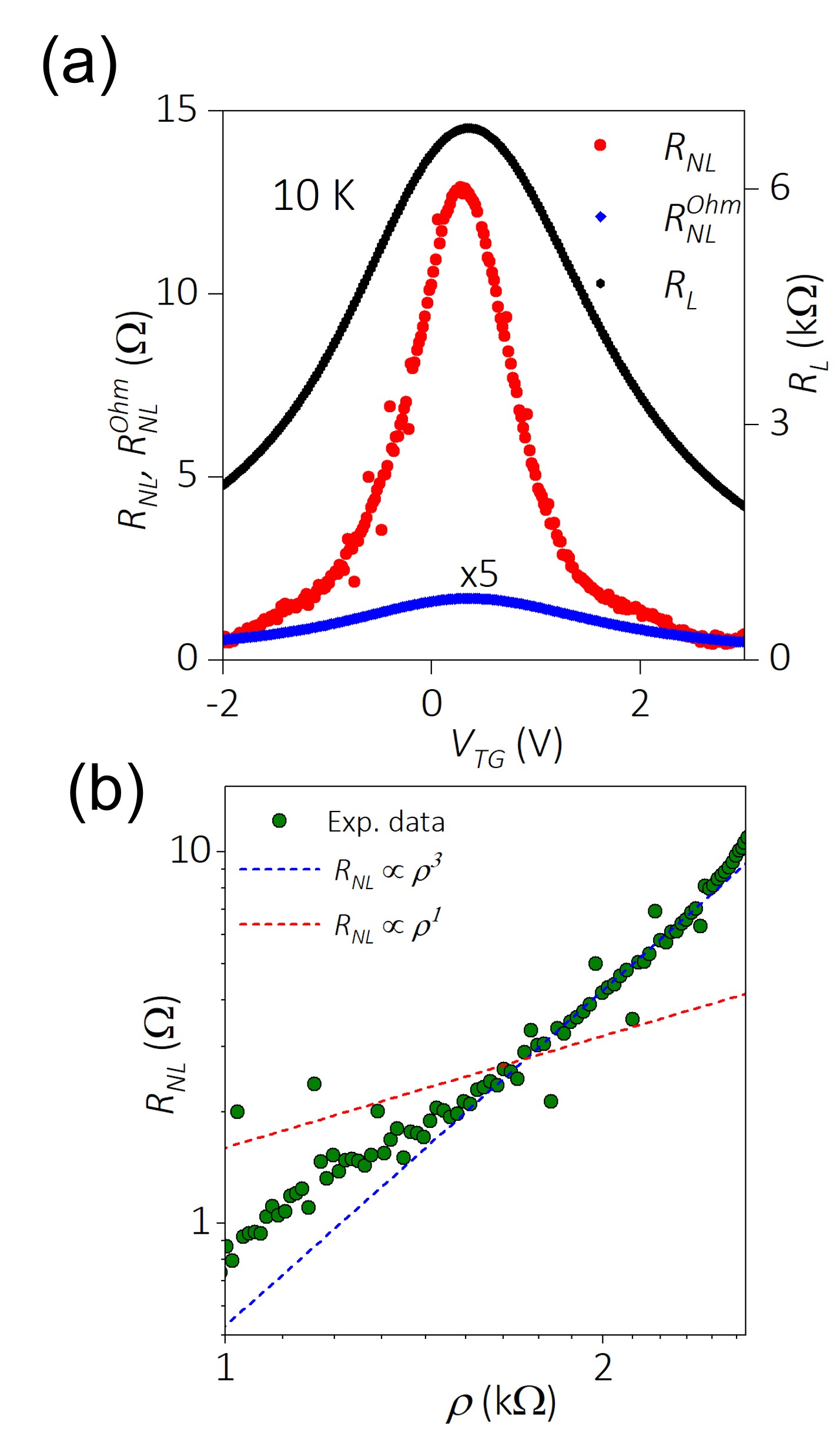}
\centering
\caption{\label{fig:two} (a) Measured local ($R_{\rm{L}}$) and non-local ($R_{\rm{NL}}$) resistance for the heterostructure at 10K. The blue line is the calculated Ohmic contribution to the non-local resistance. (b) $R_{\rm{NL}}$ and $\rho$ follows a cubic scaling relation ($R_{\rm{NL}} \propto \rho^{3}$) indicating that the measured $R_{\rm{NL}}$ originates from VHE.}
\end{figure*}

 The local and non-local electrical measurements are performed following the standard four-terminal method using \small{KEITHLEY} 4200 semiconductor parameter analyzer with a high input impedance (10$^{13}$ $\ohm$) at the voltage terminals. A separate source meter (\small{KEITHLEY} 2400) is used to apply gate voltage. Figure \ref{fig:two}a compares the local ($R_{\rm{L}}$) and non-local ($R_{\rm{NL}}$) resistance measurement results for Device A at 10 K. For $R_{\rm{L}}$, a current is applied between terminal 2 and 3 and the voltage drop between terminals 9 and 8 is detected giving $R_{\rm{L}}$ = $V_{\rm{9,8}}/I_{\rm{2,3}}$. For $R_{\rm{NL}}$ measurement, the current is applied at the local terminals 3 and 8, and the voltage drop at terminals 2 and 9 is measured, giving $R_{\rm{NL}}$ = $V_{\rm{2,9}}/I_{\rm{3,8}}$. The length and width of the Hall bar in the measured region are 2.5 $\mu$m and 1 $\mu$m, respectively. A strong $R_{\rm{NL}}$ signal is detected around the charge neutrality point (CNP) with zero electric or magnetic field applied across the layers. The peak of the $R_{\rm{NL}}$ generally appears at the CNP. The shift in the $R_{\rm{NL}}$ peak is attributed to the in-homogeneity in the bilayer graphene channel, especially since the graphene is on $\rm{SiO}_{2}$ substrate. To rule out the possibility of diffusive charge contribution to the measured non-local signal, we also calculated the Ohmic contribution using the formula  $R_{\rm{NL}}^{Ohm} = R_{\rm{L}} \left(\frac{W}{\pi L} \right){\rm{exp}} \left(- \frac{\pi L}{W} \right)$ \cite{Abanin-2011}. The calculated Ohmic contribution is at least one order of magnitude less than that of the measured $R_{\rm{NL}}$, thereby ruling out the possibility of diffusive charge contribution. One possible origin of the observed $R_{\rm{NL}}$ would be the VHE. The Berry curvature induced VHE and the resultant transverse valley Hall conductivity ($\sigma_{xy}^{\rm{VH}}$) is related to the measured $R_{\rm{NL}}$ as

\begin{equation}
R_{\rm{NL}} = \frac{1}{2} \left( \frac{\sigma_{xy}^{\rm{VH}}}{\sigma_{xx}} \right) ^{2}
                \frac{W}{\sigma_{xx} l_{\rm{v}}}
                {\rm{exp}} \left( -\frac{L}{l_{\rm{v}}} \right)
\label{eq:one}
\end{equation}

where $L$ and $W$ are the length and width of the device, $l_{\rm{v}}$ is the valley diffusion length and $\sigma_{xx} = 1/\rho$ is the conductivity. In the small valley Hall angle regime ($\sigma_{xy}^{\rm{VH}}/ \sigma_{xx}) \ll 1$), $R_{\rm{NL}}$ and $\rho$ holds a cubic scaling relation ($R_{\rm{NL}} \propto \rho^{3}$). Thus we plotted $R_{\rm{NL}}$ as a function of $\rho$ as shown in figure \ref{fig:two}b, which exhibits a clear cubic relation implying that the measured $R_{\rm{NL}}$ indeed originates from the VHE. Here, $\rho$ is defined as $R_{\rm{L}}(W/L)$ where $W$ and $L$ are the width and length of the measured part of the channel, respectively. 

\begin{figure*}[t]
\includegraphics[scale=0.6]{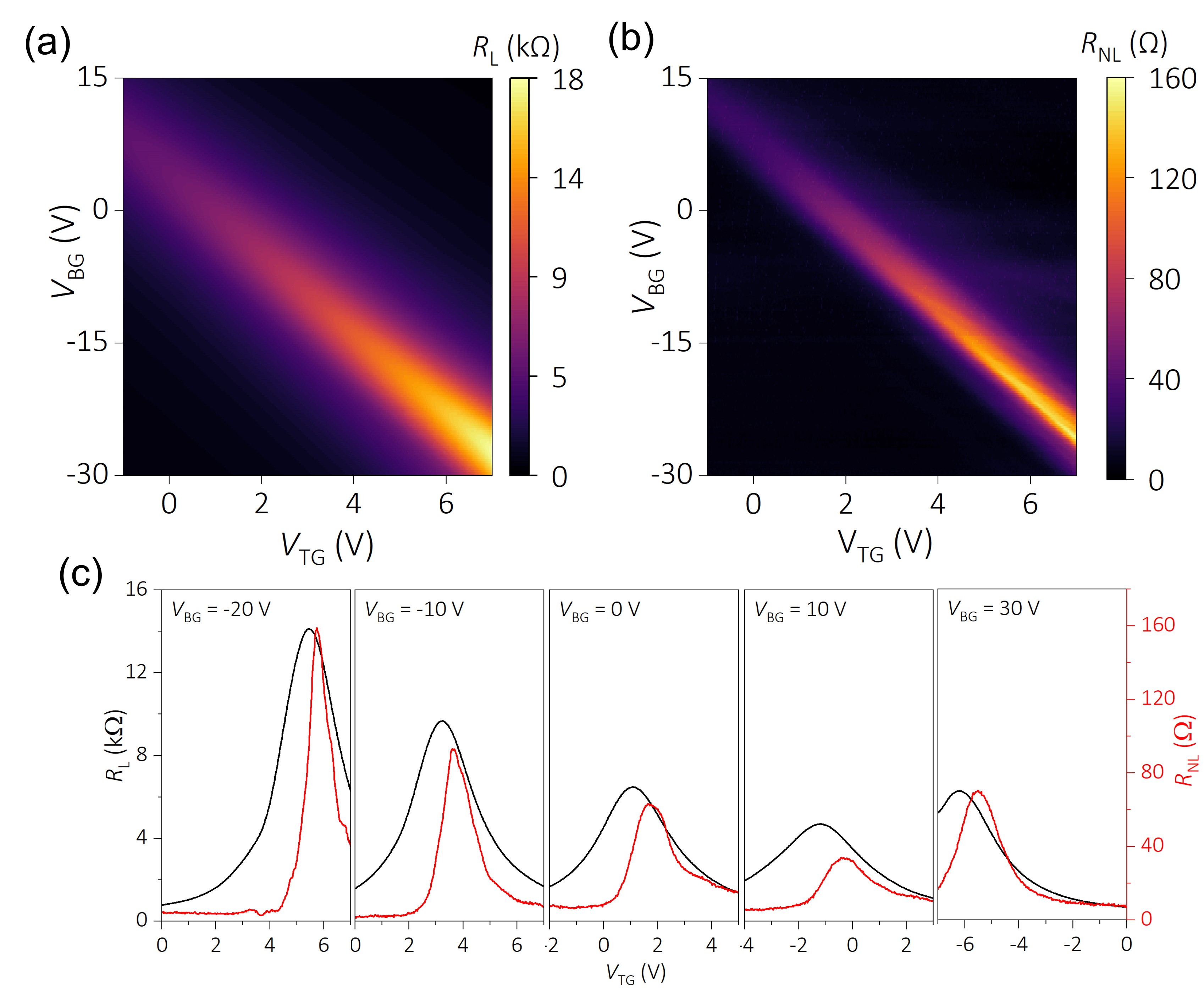}
\centering
\caption{\label{fig:three} Heat map of the (a) $R_{\rm{L}}$ and (b) $R_{\rm{NL}}$ as a function of top-gate and bottom-gate. (c) $R_{\rm{L}}$ and $R_{\rm{NL}}$ measured as a function of top-gate voltage with back-gate fixed at different values. All the measurements are performed at 10 K.}
\end{figure*}

Next, we investigate the effect of an electric displacement field applied across the bilayer graphene on both $R_{L}$ and $R_{NL}$. Figure \ref{fig:three}a and b show the heat map of the $R_{\rm{L}}$ and $R_{\rm{NL}}$ respectively as a function of both top-gate and bottom-gate voltage. It can be seen that the $R_{\rm{NL}}$ is narrower than $R_{\rm{L}}$ indicating the difference in the physical origin of both peaks. Applying voltages on the top and bottom gates allows the independent control of carrier concentration and the displacement field. The displacement fields related to the top and bottom gates ($V_{\rm{TG}}$ and $V_{\rm{BG}}$) are defined as:

\begin{equation}
\begin{split}
D_{TG} = -\epsilon_{TG} (V_{TG} - V_{TG}^{0})/d_{TG} \\ 
D_{BG} = -\epsilon_{BG} (V_{BG} - V_{BG}^{0})/d_{BG}
\label{eq:two}
\end{split}
\end{equation}

where $\epsilon_{TG}(\epsilon_{BG})$ and $d_{TG}(d_{BG})$ are the dielectric constant and thickness of the top(bottom) layer. $V_{TG,BG}^{0}$ is the voltage offset, which is -1.2 V and 10 V respectively, for Device A. The difference between the two displacement fields gives the carrier doping, and the average of the two is the net displacement fields. Figure \ref{fig:three}c shows the evolution of $R_{\rm{L}}$ and $R_{\rm{NL}}$ as a function of the displacement field. Here $R_{\rm{L}}$ and $R_{\rm{NL}}$ are plotted as a function of top gate voltage with the back gate fixed at different values. As mentioned earlier, the heterostructure at the pristine state ($V_{\rm{BG}} = 0V$) shows a clear non-local signal, suggesting an in-built asymmetry present in the heterostructure. Application of a negative electric field ($V_{\rm{BG}}$ = -20V and -10V) increases the intensity of both $R_{\rm{L}}$ and $R_{\rm{NL}}$. This implies that a negative electric field widens the band gap and enhances the asymmetry between the layers. Whereas the application of a small positive electric field ($V_{\rm{BG}} = 10V$) reduces the intensity of both peaks, suggesting a reduction in the band gap and the asymmetry. However, the application of a strong positive electric field ($V_{\rm{BG}} = 30V$) yet again increases the intensity of both $R_{\rm{L}}$ and $R_{\rm{NL}}$, implying an increase in band gap and asymmetry of the heterostructure. 

\begin{figure*}[h]
\includegraphics[scale=0.9]{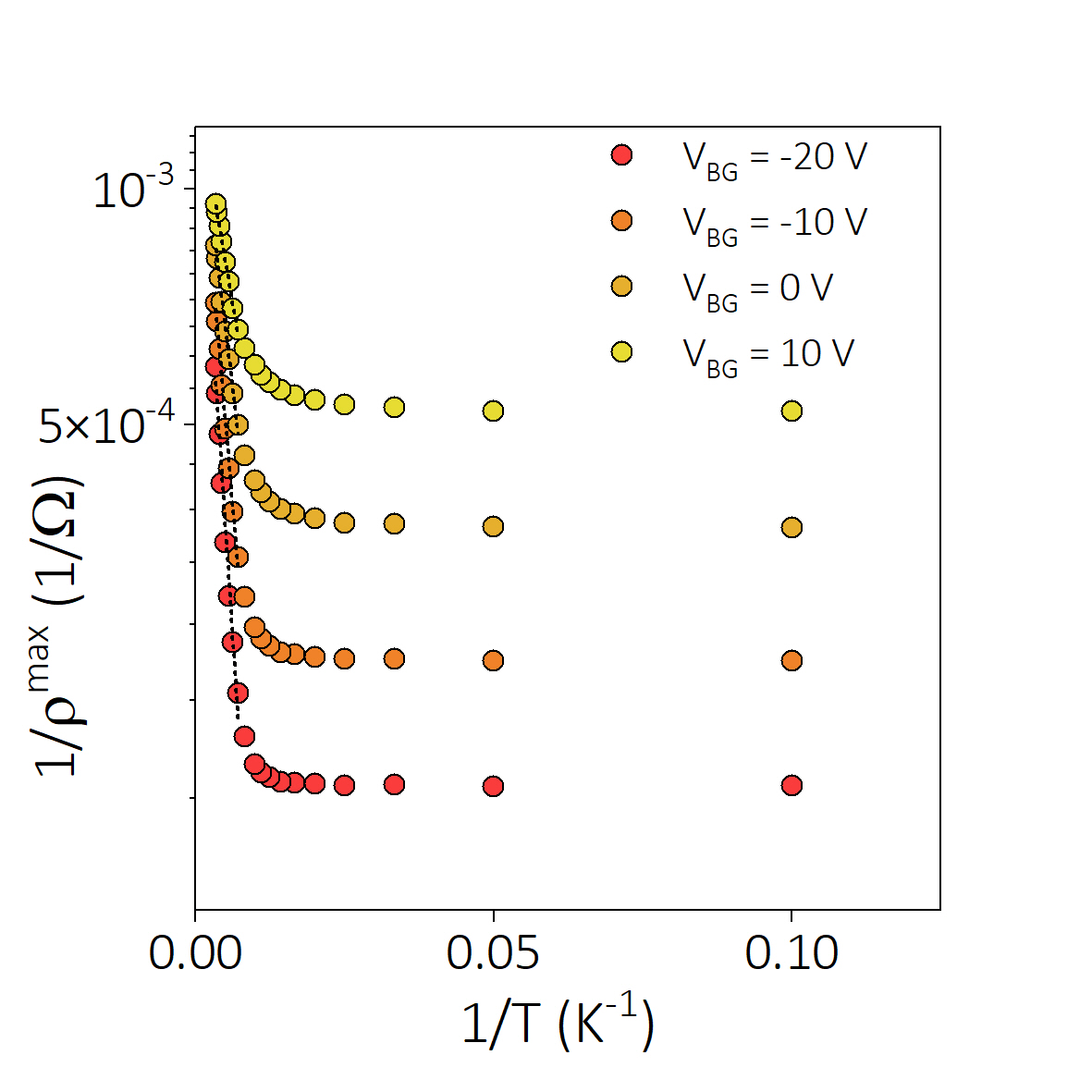}
\centering
\caption{\label{fig:four}  $1/ \rho_{\rm{L}}^{\rm{max}}$ as a function of $1/T$ plot for different $V_{\rm{BG}}$ showing the temperature dependence of $\rho_{\rm{L}}^{\rm{max}}$. The dotted line indicates the fitting to equation \ref{eq:three} in the high-temperature regime from which the band gap of the heterostructure at different $V_{\rm{BG}}$ is extracted.}
\end{figure*}

To validate the above hypothesis, we measured the temperature dependence of the $R_{\rm{L}}$ at different gate voltages to calculate the band gap. The maximum of the local resistivity ($\rho_{\rm{L}}^{\rm{max}}$) at the high-temperature regime is related to the band gap as:

\begin{equation}
\frac{1}{\rho_{L}^{\rm{max}}} = \frac{1}{\rho_{L}}
                              {\rm{exp}} \left( -\frac{E^{L}}{k_{\rm{B}} T} \right)
\label{eq:three}
\end{equation}

where $\rho_{L}$ is the local resistivity, $E^{L}$ is the activation energy, $k_{\rm{B}}$ is the Boltzmann constant and $T$ is the temperature. The band gap $E_{\rm{g}}$, defined as 2$E^{\rm{L}}$, can be extracted by plotting $1/ \rho_{\rm{L}}^{\rm{max}}$ as a function of $1/T$ as shown in figure \ref{fig:four}. The dotted line is the fit to equation \ref{eq:three} at the high-temperature regime. Table \ref{tab:table1} shows the extracted band gap values at different $V_{\rm{BG}}$. The heterostructure has an intrinsic band gap of 25 meV (at $V_{\rm{BG}}$ = 0 V). Application of a negative displacement field enhances the band gap (35.3 meV at $V_{\rm{BG}}$= -10 V and 45.5 meV at $V_{\rm{BG}}$= -20 V). At the same time, the application of a positive displacement field reduces the band gap (17.5 meV at $V_{\rm{BG}}$= 10 V). This is consistent with the earlier observation of increase(decrease) in the $R_{\rm{L}}$ and $R_{\rm{NL}}$ peak intensity at negative(positive) displacement field. 

\begin{table}
\caption{\label{tab:table1}Band gap extracted from the $1/ \rho_{\rm{L}}^{\rm{max}}$ versus $1/T$ plot for different $V_{\rm{BG}}$. $E_{\perp}$ is the electric field corresponding to each $V_{\rm{BG}}$ at $\rho_{\rm{L}}^{\rm{max}}$.}
\begin{ruledtabular}
\begin{tabular}{lccc}
$V_{\rm{BG}}$ (V)  & $E_{\perp}$ (V/nm)  & Band gap (meV) & $R_{NL}^{\rm{max}} (\Omega)$ \\
\hline
-20 & -0.44 & 45.5 & 160  \\
-10 & -0.30 & 35.3  & 94 \\
0 & 0.14 & 24.8 & 64  \\
10 & 0.00 & 17.5 & 34  \\

\end{tabular}
\end{ruledtabular}
\end{table}

\begin{figure*}[h]
\includegraphics[scale=0.5]{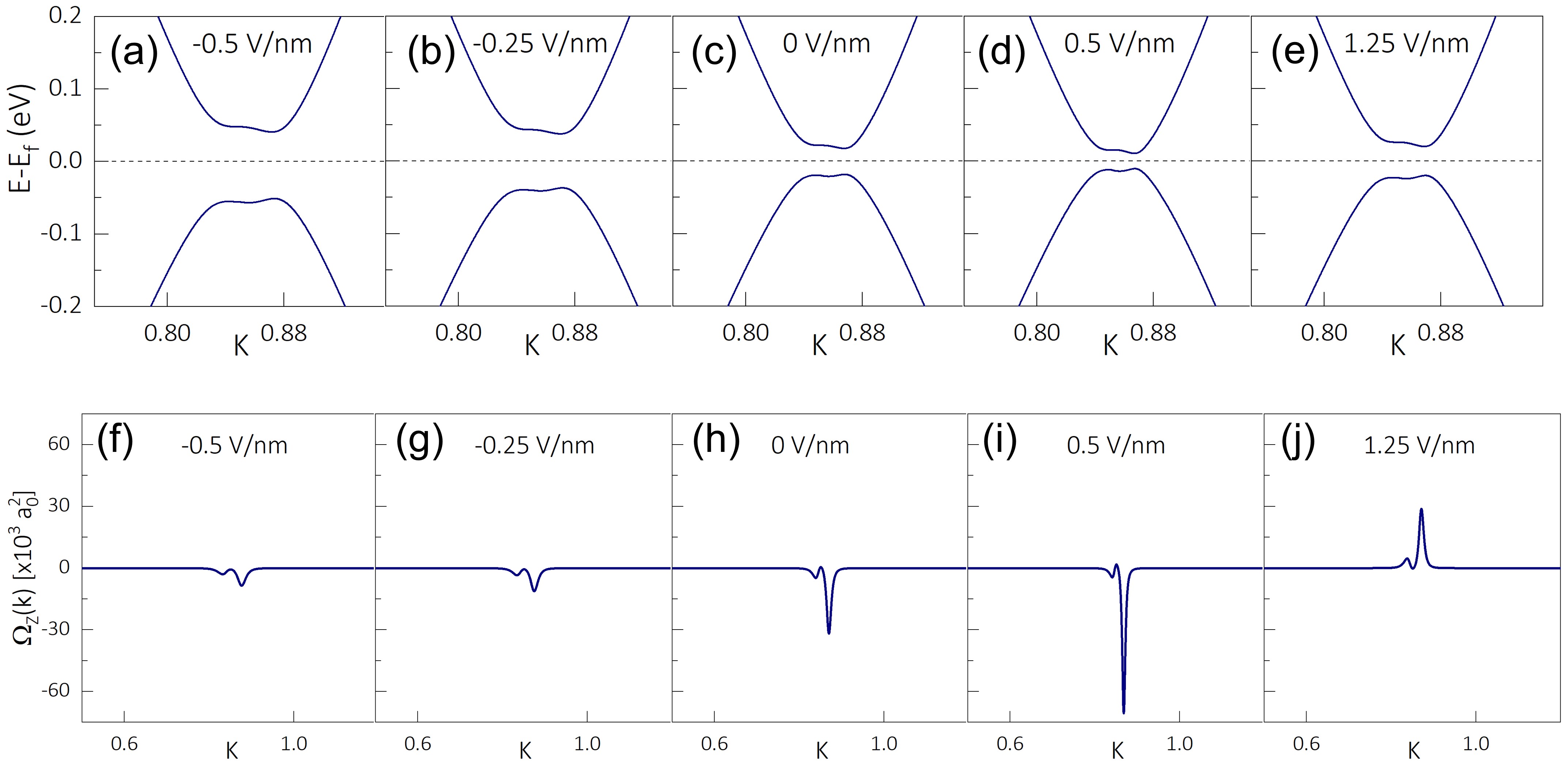}
\centering
\caption{\label{fig:five} Electronic band structure calculated for the hBN/bilayer graphene heterostructure at electric fields of magnitude (a) -0.5 V/nm, (b) -0.25 V/nm, (c) 0 V/nm, (d) 0.5 V/nm and (e) 1.25 V/nm. The path of the band structure calculation is $M$ $\rightarrow$ $K$ $\rightarrow$ $\Gamma$. Berry curvature calculated at $K$ high symmetry point for the heterostructure at electric fields of magnitude (f) -0.5 V/nm, (g) -0.25 V/nm, (h) 0 V/nm, (i) 0.5 V/nm and (j) 1.25 V/nm. The path of the Berry curvature calculation is the same as the band structure calculation.}
\end{figure*}

We have also performed \textit{ab initio} calculations to substantiate the experimental observations (see \cite{dft} for calculation details). Figure \ref{fig:five}a-e shows the electronic band structure calculated for the hBN/bilayer graphene heterostructure at different electric fields applied across the layers. The heterostructure has an intrinsic band gap of 36 meV (Fig. \ref{fig:five}c), implying asymmetry between the layers. As the low energy bands are constituted by the non-dimer atoms in the bottom and top layers of the bilayer graphene, the hBN induces different potentials between them, which opens a band gap. Applying a negative electric field (Fig. \ref{fig:five}a-b) introduces additional asymmetry between the layers, enhancing the band gap (75 meV for -0.25 V/nm and 92 meV for -0.5 V/nm). However, applying a positive electric field initially works against the inbuilt asymmetry between the layers, reducing the band gap to 21 meV (Fig. \ref{fig:five}d). At higher positive electric fields, the electric potential surpasses the intrinsic potential difference between the layers and widens the band gap further (40 meV for 1.25 V/nm), as shown in figure \ref{fig:five}e. The results of the band structure calculation strongly agree with the experimental observations. Next, we look at the Berry curvature calculated for the heterostructure. Berry curvature for an electronic band is defined as

\begin{equation}
\mathbf{\Omega}_{n} (\mathbf{k}) = i  \frac{{\hbar}^{2}}{m^{2}}\sum_{n \neq n^{'}}
\frac{\braket{u_{n,\mathbf{k}}|\hat{\mathbf{p}}|u_{n^{'},\mathbf{k}}}
\times 
\braket{u_{n^{'},\mathbf{k}}|\hat{\mathbf{p}}|u_{n,\mathbf{k}}}}
{(\varepsilon_{n} - \varepsilon_{n^{'}})^{2}}
\label{eq:four}
\end{equation}

where $\ket{u_{n,\boldsymbol k}}$ is the periodic part of the Bloch function, $\hat{\mathbf{p}}$ is the momentum operator, $\varepsilon_{n}$ is the energy of the $n$th band and $\varepsilon_{n^{'}}$ represents the energy of all other bands. The total Berry curvature is the sum of the individual occupied band's Berry curvature ($\boldsymbol\Omega(\textbf{k}) = \sum_{n} f_{n} \boldsymbol\Omega_{n}(\textbf{k})$). The Wannier interpolation scheme dictates that a pair of bands that are either occupied or unoccupied have a negligible contribution to the total Berry curvature \cite{X-Wang-2006}. The major contribution to the total Berry curvature comes from a pair of bands where one is occupied and another unoccupied, such as the low energy bands in the hBN/bilayer graphene heterostructure. In addition, the denominator of equation \ref{eq:four} suggests that Berry curvature value varies as the square of the energy difference between two adjacent bands. Thus in our case, the Berry curvature changes with the electric field as the band gap changes. The heterostructure has an intrinsic non-zero Berry curvature as shown in figure \ref{fig:five}h. A negative electric field reduces the magnitude of the Berry curvature as it widens the gap between the low energy bands (Fig. \ref{fig:five}f-g). On the other hand, a small positive electric field (0.5 V/nm) enhances the magnitude of the Berry curvature due to the narrow band gap (Fig. \ref{fig:five}i). However, two key differences could be observed at a higher positive electric field (Fig. \ref{fig:five}j). One, the magnitude of the Berry curvature reduces owing to the widening of the band gap. The second is the change in the polarity of the Berry curvature, which implies that the polarity of the layer asymmetry switches direction at higher positive electric fields. 

In conclusion, we have observed Berry curvature induced VHE in non-encapsulated hBN/bilayer graphene heterostructure where the hBN and bilayer graphene are aligned with near-zero twist angle. The VHE is detected as a non-local resistance near the CNP. The cubic relation observed between $R_{\rm{NL}}$ and $\rho$ validates that the measured $R_{\rm{NL}}$ indeed originates from the VHE. The measured $R_{\rm{NL}}$ could be manipulated with the application of a displacement field across the layers, which is attributed to the change in the electronic band structure and the asymmetry of the bilayer graphene under a displacement field. The intrinsic band gap of the heterostructure and its evolution under the displacement field is confirmed from the temperature-dependent $R_{\rm{L}}$ measurement in the high-temperature regime. The experimental observations were substantiated with \textit{ab initio} calculations which showed that the heterostructure has an intrinsic band gap and a non-zero Berry curvature, both of which could be controlled by a perpendicular electric field. 

\textbf{Acknowledgements:} Part of this research was supported by Toshiba electronic devices \& storage corporation's academic encouragement program.

\pagebreak

\widetext
\begin{center}
\hfill \break
\hfill \break
\textbf{\Large Supplementary Information}
\end{center}
\setcounter{equation}{0}
\setcounter{figure}{0}
\setcounter{table}{0}
\setcounter{page}{1}
\makeatletter
\renewcommand{\theequation}{S\arabic{equation}}
\renewcommand{\thefigure}{S\arabic{figure}}
\renewcommand{\bibnumfmt}[1]{[S#1]}
\renewcommand{\citenumfont}[1]{S#1}

\section{Results obtained from Device B}

\textbf{ I. Optical image}
\hfill \break

\begin{figure*}[h]
\includegraphics[scale=0.4]{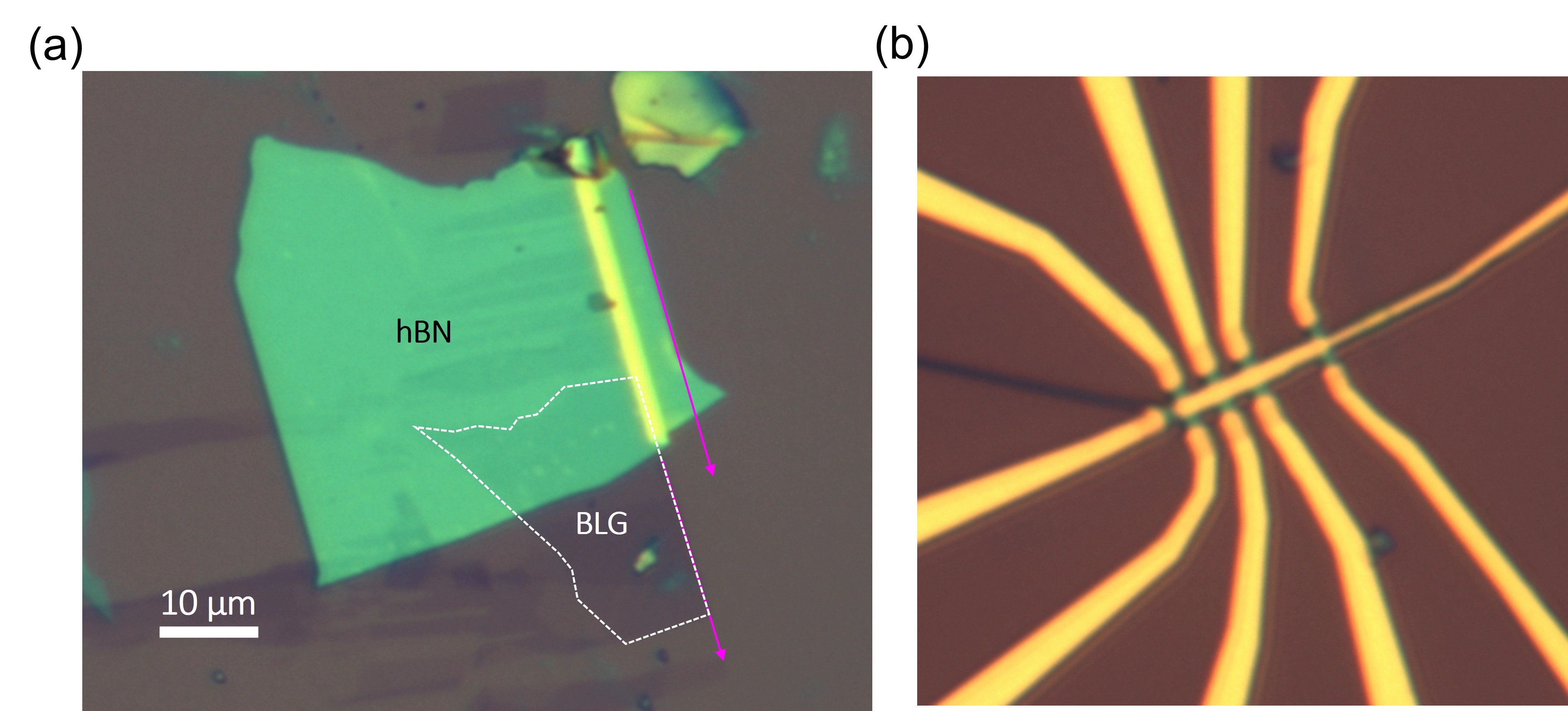}
\centering
\caption{\label{fig:s-one} (a) Optical image of Device B. The dotted line shows the graphene region. The arrow indicates the edges of graphene and hBN, which are aligned. (b) Optical image of the final device after fabricating contacts.}
\end{figure*}

Figure \ref{fig:s-one}a shows the optical image of Device B where hBN and bilayer graphene edges are aligned with near-zero twist angle. The dotted line shows the graphene region. The arrow indicates the edges of graphene and hBN, which are aligned. Figure \ref{fig:s-one}b is the optical image of the final device. A leakage current was observed from the top gate on this device. This could be due to the insufficient thickness of the hBN layer. Thus no displacement field measurement was performed on this device.

\textbf{ II. Gate characteristics}
\hfill \break

\begin{figure*}[h]
\includegraphics[scale=0.6]{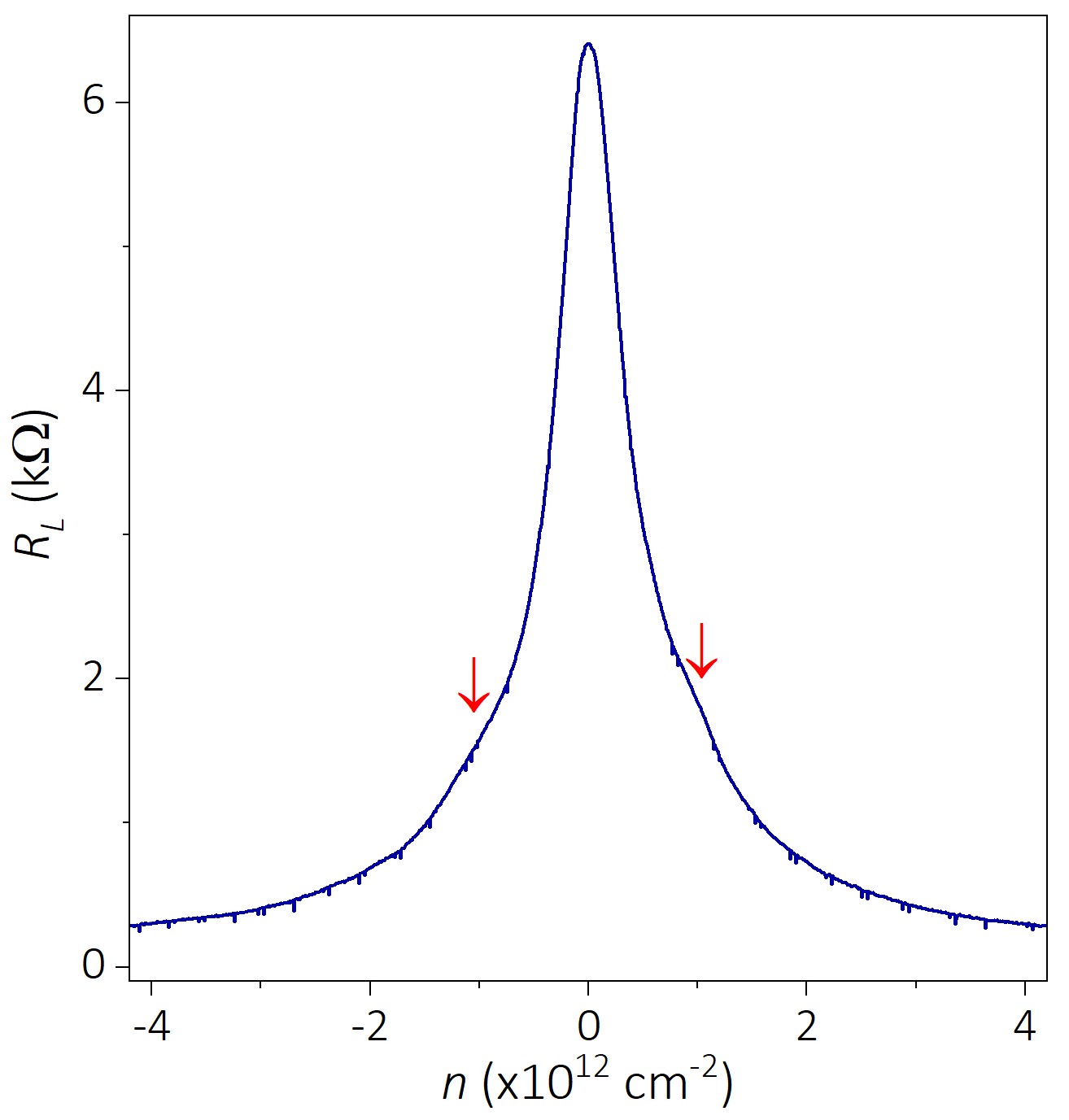}
\centering
\caption{\label{fig:s-two} Gate characteristics of Device B.Besides the MDP, two SDPs (indicated by arrows) can be seen on both sides of the MDP.}
\end{figure*}

Figure \ref{fig:s-two} shows the gate characteristics of Device B. The extracted field-effect carrier mobility from the linear region around the MDP is $\approx$ 9 400 cm$^2$ V$^{-1}$ s$^{-1}$ and $\approx$ 7 100 cm$^2$ V$^{-1}$ s$^{-1}$ respectively for the holes and electrons. In addition to the MDP, two SDPs, indicated by arrows, can be seen on both sides of the MDP, affirming the formation of a moire superlattice between hBN and bilayer graphene. The SDPs appear very close to the MDP at a carrier density of $\pm$ 1.05 $\times$ 10$^{12}$ cm$^{-2}$. This corresponds to a moire wavelength of 20 nm, much higher than the theoretically predicted maximum value of 13.8 nm. The moire wavelength $\lambda$ is derived as \cite{Yankowitz2-2012}

\begin{equation}
\lambda = \frac{(1+\delta)a}{\sqrt{2(1+\delta)(1-cos\phi) + \delta^{2}}}
\label{eq:s-one}
\end{equation}

where $\delta$ is the lattice mismatch between graphene and hBN, $a$ is the graphene lattice constant (0.246 nm), $\phi$ is the rotation angle between graphene and hBN. Considering that the rotation angle is 0, a moire wavelength of 20 nm corresponds to a lattice mismatch of 1.25\%, which is smaller than the lattice mismatch between pristine hBN and graphene (1.79\%). This implies that the graphene is stretched. We attribute the stretching in graphene to the non-uniform surface of SiO$_{2}$ as the bilayer graphene is in direct contact with it.

\hfill \break

\textbf{ III. Local and non-local measurement results}
\hfill \break

\begin{figure*}[h]
\includegraphics[scale=0.7]{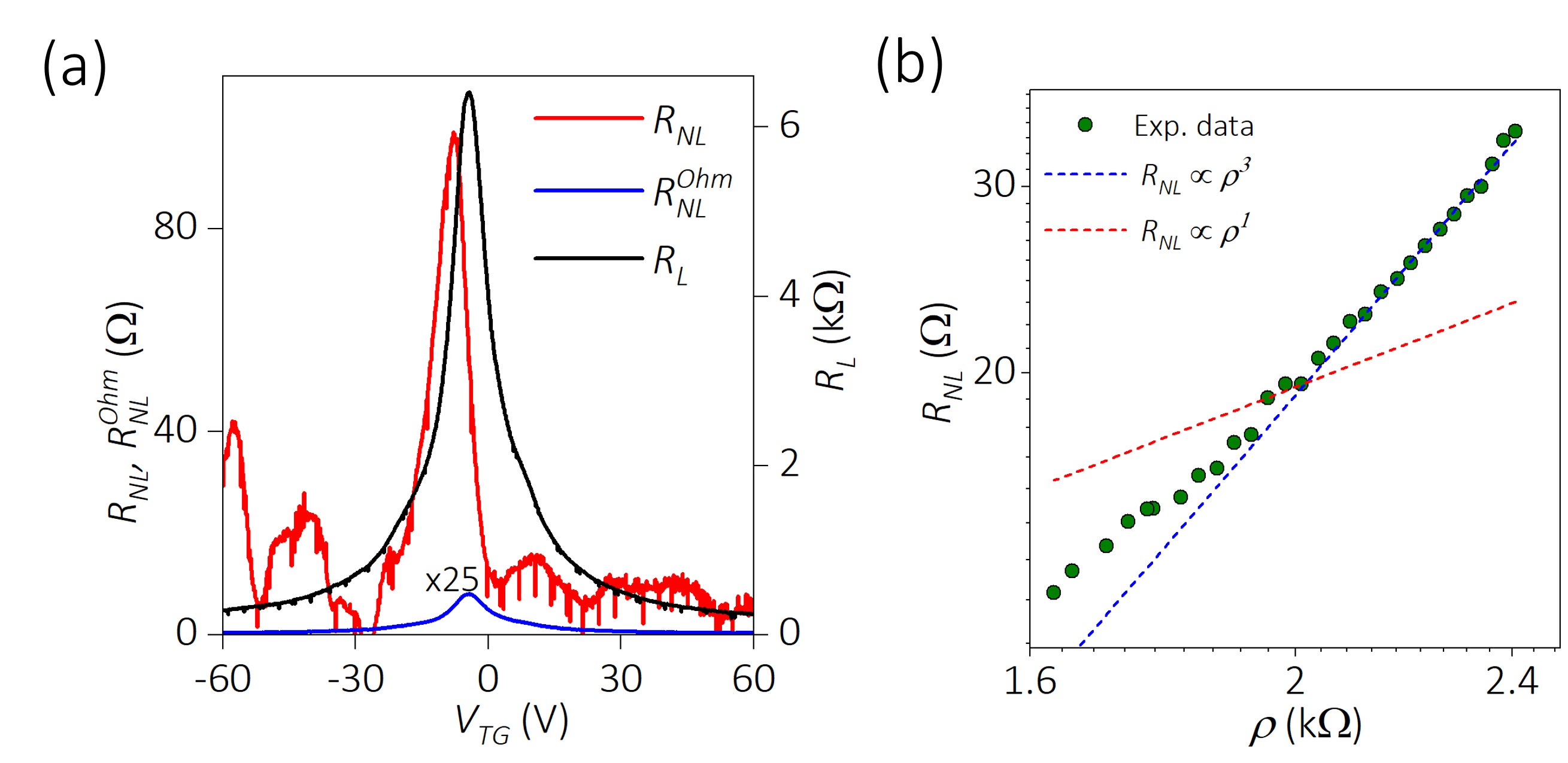}
\centering
\caption{\label{fig:s-three} (a) Measured ($R_{\rm{L}}$) and ($R_{\rm{NL}}$) for Device B at 10K. The blue line is the calculated Ohmic contribution (multiplied by 25) to the non-local resistance. (b) Cubic scaling relation between $R_{\rm{NL}}$ and $\rho$.}
\end{figure*}

Figure \ref{fig:s-three}a shows the local ($R_{\rm{L}}$) and non-local ($R_{\rm{NL}}$) resistance for Device B at 10K. The calculated Ohmic contribution to the $R_{\rm{NL}}$ is negligible compared to the magnitude of the $R_{\rm{NL}}$. Figure \ref{fig:s-three}b is the cubic scaling relation plotted between $R_{\rm{NL}}$ and $\rho$, where $\rho$ is defined as $R_{\rm{L}}(W/L)$. The width $W$ and length $L$ of the measured part of the channel are 1 $\mu$m and 2.5 $\mu$m, respectively. The clear cubic relation between $R_{\rm{NL}}$ and $\rho$ implies that the non-local signal originates from the valley Hall effect.

\end{document}